\documentclass[notitlepage, 10pt]{article}

\usepackage[top=2cm, bottom=2cm, left=1.25cm, right=1.25cm]{geometry}
\usepackage{graphicx}

\usepackage{titlesec}
\usepackage{sectsty}
\usepackage{authblk}
\usepackage{siunitx}
\usepackage{amsmath}
\usepackage[super,comma,sort&compress]{natbib}
\titlespacing\section{0pt}{2pt plus 4pt minus 2pt}{0pt plus 2pt minus 2pt}
\sectionfont{\fontsize{12}{15}\selectfont}
\paragraphfont{\fontsize{9}{9}\selectfont}
\usepackage{braket}
\usepackage[font=small,labelfont=bf]{caption}
\usepackage{lineno}
\usepackage{nameref}
\usepackage{newfloat}
\usepackage[expansion=true]{microtype}

\DeclareFloatingEnvironment{metfig}

\usepackage{hyperref}
\hypersetup{
	colorlinks=true,
	linkcolor=black,
	citecolor=black,
	filecolor=black,
	urlcolor=black,
}

\title{Single-electron spin resonance in a nanoelectronic device using a global field}
\date{}
\author[1,*]{E. Vahapoglu}
\author[1,*]{J. P. Slack-Smith}
\author[1]{R. C. C. Leon}
\author[1]{W. H. Lim}
\author[1]{F. E. Hudson}
\author[1]{T. Day}
\author[1]{T. Tanttu}
\author[1]{C. H. Yang}
\author[1]{A. Laucht}
\author[1,$\dagger$]{A. S. Dzurak}
\author[1,$\dagger$]{J. J. Pla}
\affil[1]{School of Electrical Engineering and Telecommunications, UNSW Sydney, Sydney, NSW 2052, Australia.}
\affil[*,$\dagger$]{These authors contributed equally to this work}

\begin{document} 


\maketitle 

\begin{abstract}
\textbf{Spin-based silicon quantum electronic circuits offer a scalable platform for quantum computation, combining the manufacturability of semiconductor devices with the long coherence times afforded by spins in silicon. Advancing from current few-qubit devices to silicon quantum processors with upwards of a million qubits, as required for fault-tolerant operation, presents several unique challenges, one of the most demanding being the ability to deliver microwave signals for large-scale qubit control. Here we demonstrate a potential solution to this problem by using a three-dimensional dielectric resonator to broadcast a global microwave signal across a quantum nanoelectronic circuit. Critically, this technique utilizes only a single microwave source and is capable of delivering control signals to millions of qubits simultaneously. We show that the global field can be used to perform spin resonance of single electrons confined in a silicon double quantum dot device, establishing the feasibility of this approach for scalable spin qubit control.}
\end{abstract}


\twocolumn

\section*{Introduction}
The ability to engineer quantum systems is expected to enable a range of transformational technologies including quantum-secured communication networks, enhanced sensors and quantum computers, with applications spanning a diverse range of industries. Quantum computers are poised to significantly outperform their classical counterparts in many important problems like quantum simulation (aiding materials and drug development) and optimization. Whilst some applications are expected to be executable on medium-scale quantum computers (with 100-1000 qubits) that do not employ error correction protocols \cite{Preskill2018}, arguably the most disruptive algorithms \cite{Shor1994} will require a large-scale and fully fault-tolerant quantum computer with upwards of a million qubits \cite{Fowler2012, Lekitsch2017}. 

Of the possible physical implementations of qubits, electron spins in gate-defined silicon quantum dots stand out for their long coherence times \cite{Veldhorst2014}, their ability to operate at temperatures above 1\,K \cite{Yang2020, Petit2020} and for the potential to leverage the experience of the semiconductor industry in fabricating devices at scale. The building blocks for a spin-based quantum computer in silicon, namely high-fidelity single \cite{Veldhorst2014} and two \cite{Veldhorst2015, Huang2019, Xue2019} qubit gates, have been demonstrated and attention is now being directed towards the engineering challenges of constructing a large-scale processor \cite{Gonzalez2020}. To achieve this ambitious goal, significant hurdles associated with qubit measurement and control must be overcome, such as how to deliver the microwave control signals to many qubits simultaneously, without disturbing the fragile cryogenic environment in which the processor operates.

One approach for spin qubit control successfully deployed in current few-qubit devices is based on a direct magnetic drive using an on-chip transmission line (TL) \cite{Dehollain2013}. A strong microwave current is passed through a wire placed close to the quantum dot to generate an alternating magnetic field. Localization of the control field to the wire demands a number of transmission lines that scales with the total number of qubits \cite{Li2018}, while multiple high-frequency coaxial lines would be needed to deliver the control signals into the cryogenic system. The large microwave currents running through the processor in such a scheme raises concerns of heating, whilst the control lines themselves take up valuable chip real-estate.

Electric dipole spin resonance (EDSR) \cite{Pioro2008, Kawakami2014, Takeda2016, Watson2018, Yang2020, Leon2020} is an alternative method for producing local spin control. For electron spin qubits, this technique simulates a magnetic drive by exploiting magnetic field gradients produced by nanomagnets in conjunction with microwave electric fields applied directly to the qubit gate electrodes. Integration of EDSR in large 2D qubit arrays poses several challenges. In addition to the required transversal gradients, stray longitudinal gradients from the nanomagnets can cause spin decoherence. Furthermore, as with the TL-based direct magnetic drive, this technique requires microwave signals to be applied across many control lines, with multiple coaxial cables entering the system and both cross-talk and layout issues to overcome.

\begin{figure*}[]
	\centering
	\includegraphics[width=1\linewidth]{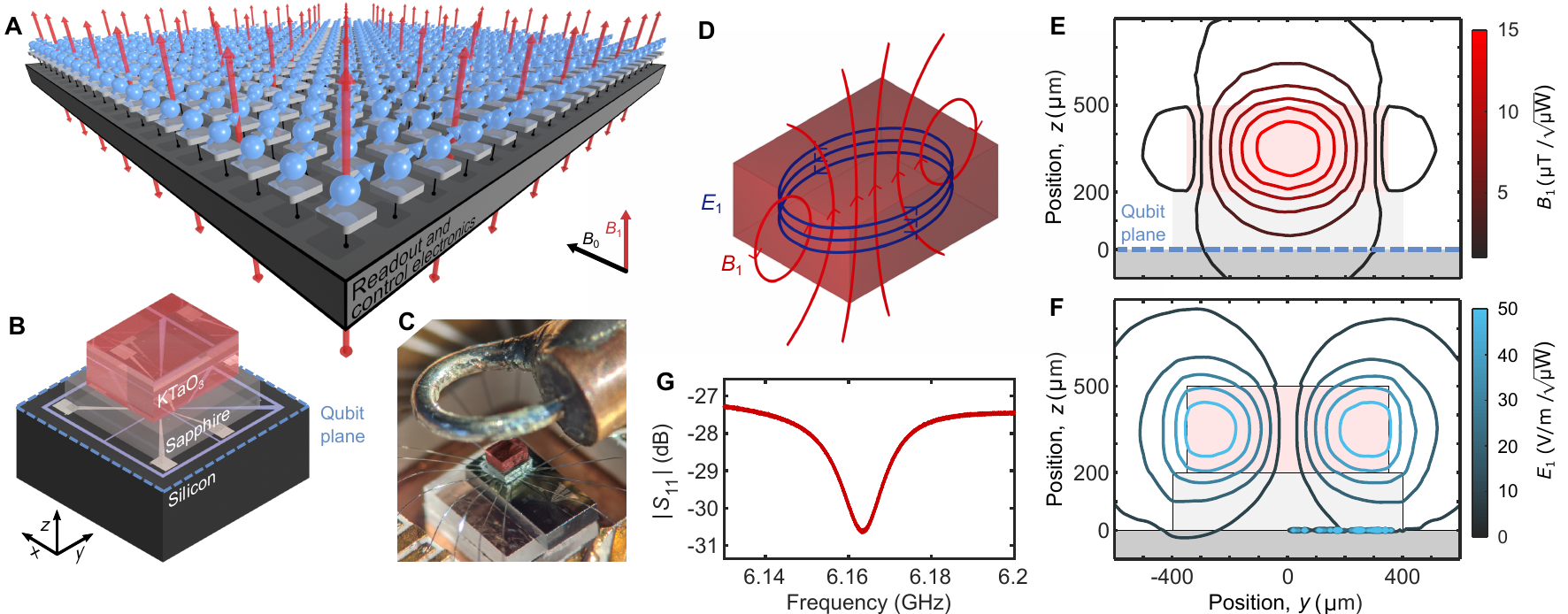}
	\caption{\textbf{Device stack for off-chip ESR and dielectric resonator simulations.}
		\textbf{(A)} A 3D render of the vision for large-scale qubit control using a global microwave field.
		\textbf{(B)} 3D graphic of the global control device stack as used in our experiments, including silicon quantum nanoelectronic device (bottom), sapphire dielectric spacer (middle) and potassium tantalate (KTaO$_3$) dielectric microwave resonator (top). The DR is a $0.7$\, $\times$ $0.55$\, $\times$ $0.3$\,mm$^3$ rectangular prism.
		\textbf{(C)} A photograph of the device and coaxial loop coupler.
		\textbf{(D)} Schematic depicting directions of the magnetic and electric field components of the TE$_{11\delta}$ mode in the dielectric resonator.
		\textbf{(E-F)} Finite-element simulation of the magnetic field $z$-component magnitude (E) and electric field magnitude (F) of the TE$_{11\delta}$ mode supported by the DR.
		\textbf{(G)} $S_{11}$ reflection parameter measurement near the fundamental mode of the dielectric resonator, measured via the coaxial loop coupler.
	}
	\label{fig:1}
\end{figure*}

An elegant and potentially scalable spin qubit control solution was anticipated in the 1998 Kane proposal for a silicon quantum computer \cite{Kane1998}. In this approach a single uniform global magnetic field is radiated across the entire processor \cite{Veldhorst2017, Hill2015} (see Fig.~\ref{fig:1}A). Qubit operations are realized by applying potentials locally to spins to bring them in and out of resonance with the global field \cite{Laucht2015}. This method utilizes a single microwave source and does not require the direct passage of strong high-frequency currents through the processor, reducing heating and simplifying the chip layout design.

Conventional electron spin resonance (ESR) spectroscopy \cite{schweiger2001} already provides a way to deliver global microwave fields to large ensembles of spins. The first-order approach to implementing global control for spin qubits in a quantum processor would be to simply place the chip inside a conventional 3D microwave cavity. However, internal device structures (such as metal gates and bond wires) adversely affect the properties of these microwave resonators \cite{Kong15}. Furthermore, the chip is exposed to large alternating electric fields in these cavities, which interfere with (or potentially damage) the sensitive qubit readout nanoelectronics.

An important metric for a microwave cavity is the power-to-field conversion factor $C$, which quantifies how well a microwave input signal is converted to the alternating magnetic field needed to drive spin rotations. The expression $B_1 = C\sqrt{P}$ relates the magnetic drive $B_1$ to the input signal power $P$ through the conversion factor $C \propto \sqrt{Q/\omega V}$, which itself depends on the quality factor $Q$ of the cavity, its frequency $\omega$ and the microwave mode volume $V$. For low conversion factors (as obtained with conventional metallic cavity resonators), substantial powers are required to drive sufficiently fast spin rotations -- powers potentially incompatible with the cryogenic environment of the quantum processor. To raise the conversion efficiency and reduce the driving power, cavities with high $Q$ and small mode volumes are desirable.

While attempts have been made to control spins in nanoelectronic devices using 3D metallic ESR resonators \cite{simovicDesignQbandLoopgap2006}, performing spin resonance with a global field has, until now, remained elusive. Here we demonstrate ESR of single spins in a silicon metal-oxide-semiconductor (SiMOS) quantum dot (QD) device by using a compact dielectric resonator (DR) placed above the chip (Fig.~\ref{fig:1}B). The DR is constructed from potassium tantalate (KTaO$_3$ or KTO), a quantum paraelectric material that exhibits an exceptionally high dielectric constant at cryogenic temperatures and hence compact microwave mode volumes. ESR control is confirmed to be resonator-driven by observing an enhancement in the mixing of the quantum dot spin states within the dielectric resonator bandwidth. This represents the first step towards the vision (see Fig.\ref{fig:1}A) of large-scale qubit control using global magnetic fields generated off-chip.

\vspace{30pt}

\section*{Results}
\paragraph*{Dielectric Resonator and Chip Assembly.}\label{section::DR}

In order to overcome the challenges presented by global control, we desire a microwave cavity that minimizes the mode volume (whilst maximizing surface area), maintains a large quality factor and produces very little electric field at the chip. In this paper we exploit a species of ESR cavity called a dielectric resonator -- a volume of dielectric material where the sudden transition between the high dielectric constant inside the material and the low dielectric constant outside of it (vacuum) leads to the formation of standing waves. DRs are often used in conventional ESR spectroscopy \cite{Blank2003} due to their high magnetic field to power conversion factors ($C$) and low intrinsic losses. Our DR is designed to operate in the TE$_{11\delta}$ mode \cite{Geifman2005}, shown in Fig.~\ref{fig:1}D. In this mode the DR acts like a magnetic dipole (i.e. a current loop). The magnetic field is strongest along the central axis (see Fig.~\ref{fig:1}E) and extends outside of the DR, while the electric field is concentrated towards the outer edges and strongly confined inside the material (Fig.~\ref{fig:1}F).

The frequency of a dielectric resonator is inversely proportional to the dielectric constant $\varepsilon_\text{r}$ and its mode volume $V$, specifically $\omega \propto 1/(\sqrt{\varepsilon_\text{r}}V^{1/3})$. Thus, for a given frequency, the mode volume can be reduced by increasing $\varepsilon_\text{r}$. In this work we exploit the perovskite material potassium tantalate. KTO is a quantum paraelectric, a material that displays ferroelectric-like properties (such as a high dielectric constant) but whose transition to the ferroelectric phase is suppressed by quantum fluctuations \cite{Gevorgian2009}. KTO has an extraordinarily large dielectric constant at cryogenic temperatures ($\varepsilon_\text{r} \approx 4,300$ for $T < 10$\,K) and exhibits very low microwave losses ($\tan\delta \sim 10^{-4}~\text{to}~10^{-5}$) permitting high internal quality factors \cite{Geyer2005}. 

The DR is cut in the shape of a rectangular prism and integrated with the silicon quantum dot device as depicted in Fig.~\ref{fig:1} (B and C). The smallest dimension of the prism is its height, as this allows us to increase the surface area over which the $B_1$ field is generated for a given DR volume (and consequently frequency). Our finite-element simulations indicate that $B_1$ has a good uniformity across the qubit plane, varying by less than 15\,\% over a $0.2 \times 0.2\,$mm$^2$ area (see Methods). We note, however, that this uniformity is far from optimized and can readily be improved through adjustments to the DR dimensions. In order to maximize the resonator $Q$ and minimize any stray electric fields reaching the device, a $200$\,\textmu m thick sapphire spacer is placed between the DR and silicon chip. The $B_1$ field used for driving spin rotations is oriented perpendicular to the qubit plane when operating in the TE$_{11\delta}$ mode. The chip containing the silicon double-quantum-dot device is wire-bonded to a printed circuit board and housed inside a copper enclosure to shield the sample and to suppress radiation losses from the DR. Microwave power is inductively coupled into the resonator through a coaxial cable terminated in a shorted loop (see Fig.~\ref{fig:1}C). The coupling strength of the DR to the coaxial port is determined primarily by the overlap between the DR and loop magnetic fields and can be controlled ex-situ by altering the position of the loop.

Experiments are performed in a dilution refrigerator at a base temperature of 50\,mK. We probe the frequency response of the resonator via an $S_{11}$ measurement through the coaxial coupler (Fig.~\ref{fig:1}G) and extract a resonant frequency of the fundamental TE$_{11\delta}$ mode of 6.163\,GHz at 50\,mK. We measured an internal quality factor of $Q_\text{i} \approx 60,000$ for the resonator in a separate cool-down without the chip, in line with the best reported values of the microwave loss-tangent for KTO \cite{Geyer2005}. In practice, we find the quality factor of the resonator to be limited by losses in the device and extract an internal quality factor of $Q_\text{i} \approx 500$. Finite-element simulations of the complete device assembly are performed to reproduce the measured DR $S_{11}$ (see Methods) and indicate an exceptionally-large conversion factor $C \approx3\,\text{\textmu T}/\sqrt{\text{\textmu W}}$.

\begin{figure*}[!t]
	\centering
	\includegraphics[width=1\linewidth]{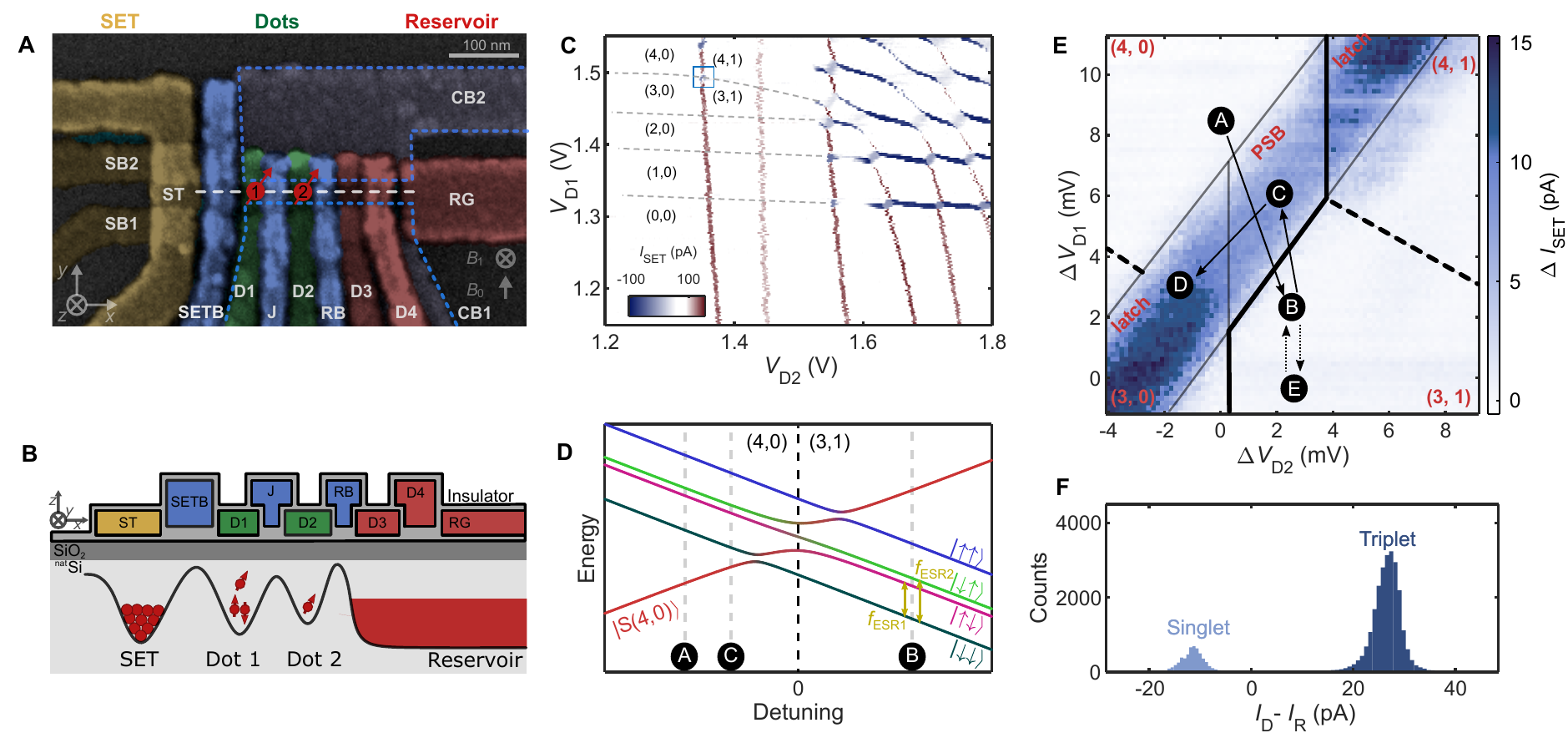}
	\caption{\textbf{Silicon double quantum dot with latched Pauli spin blockade readout.}
		\textbf{(A)} Scanning electron micrograph (SEM) of the quantum dot device with false coloring for the gate electrodes.
		\textbf{(B)} A schematic cross-section through the middle of the device (indicated with a white dashed line in panel a), showing the 3D structure of the gates and conduction band profile.
		\textbf{(C)} Charge stability diagram of the double quantum dot obtained by monitoring the current $ I_\text{SET} $ through a nearby capacitively-coupled charge sensor (a single electron transistor). The numbers in parentheses represent the charge occupancies of the double dot system Dot 1, Dot 2: ($N1$, $N2$).
		\textbf{(D)} Schematic energy diagram of the (4,0)-(3,1) anti-crossing.
		\textbf{(E)} Readout pulse sequence overlaid on a 2D color plot of the readout search signal $\Delta I_{\rm SET}$ (see Methods for details) as a function of the voltages on gates D1 and D2. The absolute voltage range is indicated by the blue rectangle in panel c. Readout is performed with the sequence B to D. The solid lines indicate transitions with high tunnel rates, the dashed lines indicate transitions with low tunnel rates, and the thin lines identify the PSB (Pauli spin blockade) and latched regions. 
		\textbf{(F)} Histogram of the latched SET current signal for $\ket{\downarrow\downarrow}$ state initialization, indicating a $\ket{\downarrow\downarrow}$ initialization infidelity of 19\,\%. 
	}
	\label{fig:2}
	
\end{figure*}

\paragraph*{Nanoelectronic Device and Spin Readout.}

The single spins in this work are provided by a \textsuperscript{nat}Si metal-oxide-semiconductor
double quantum dot device, electrostatically defined by a palladium (Pd) gate stack architecture \cite{Zhao2019}. Fig.~\ref{fig:2}A shows a scanning electron microscope (SEM) image of a device nominally identical to the one measured. A cross-section through the dot channel depicting the gate stack and conduction band profile is illustrated in Fig.~\ref{fig:2}B. Critically, instead of having a microwave transmission line \cite{Dehollain2013} or a micromagnet \cite{Leon2020} fabricated on-chip with the quantum dots, here we incorporate the coaxial loop coupler and dielectric resonator on top of the silicon device to drive spin resonance in a global field.    

To populate an arbitrary integer number of electrons ($N1$, $N2$) in Dot 1 and Dot 2, we apply positive voltages to gates D1 and D2. The electrons are loaded from an electron reservoir induced at the Si/SiO$_2$ interface by applying a positive bias to D3, D4 and RG. By lowering the voltages on confinement gates CB1, CB2 and SETB, we ensure that electrons in the two QDs are confined to small spatial regions as indicated in Fig.~\ref{fig:2}A. The barrier gates RB and J allow us to control the tunnel coupling between the dots and the reservoir (RB) as well as the interdot coupling (J). A single electron transistor (SET) nearby serves as a charge sensor to monitor the occupancy of the two QDs \cite{Veldhorst2014}.

Fig.~\ref{fig:2}C shows the charge stability diagram of the double QD system, measured via a double lock-in technique \cite{Yang2011}. The nearly horizontal lines (blue) correspond to charge transitions in Dot 1, while the nearly vertical lines (red) are caused by charge transitions in Dot 2. As we decrease $V_\text{D1}$ and $V_\text{D2}$ (the voltages applied to gates D1 and D2), electrons are depleted one-by-one from Dot 1 and Dot 2 until they are both completely empty, denoted as (0,0) in the lower left corner of the stability plot. The transitions of Dot 1 are not visible at lower $V_\text{D2}$ because the tunnel rates between Dot 1 and the reservoir for these bias conditions are smaller than the lock-in probe frequency. These transitions are indicated with gray dashed lines as a visual aid. We note that the faint vertical transition visible at $V_\text{D2}\approx1.45$\,V is due to an unintentional dot formed under gate RB. The sensing signal is weaker owing to its distance from the SET sensor.         

For the remainder of the paper, we focus on the (4,0)-(3,1) charge transition for singlet-triplet (ST) readout. Here, two of the electrons in Dot 1 form a spin-zero closed shell and do not interact with the third and fourth electrons during experiments. Spin readout is performed via the pulse sequence A to D indicated in Fig.~\ref{fig:2}E (see Methods for the experiment performed to obtain this 2D color-map). Step A prepares the double QD system in the (4,0) singlet state. Then, by ramping adiabatically from A to B, the system is initialized the triplet state $\ket{\downarrow\downarrow}$ (see the energy diagram in Fig.~\ref{fig:2}D). Pulsing from B to C attempts to move the electron from Dot 2 to Dot 1. If the electron in Dot 2 forms a singlet with the electron in Dot 1, tunneling takes place producing a change in the charge configuration from (3,1) to (4,0). However, if a triplet state is formed, Pauli spin blockade (PSB) prevents the electron from tunneling and the system remains in the (3,1) charge configuration \cite{Johnson2005}. The charge and therefore spin states can be differentiated by monitoring the SET current $I_\text{SET}$, a process referred to as spin-to-charge conversion. Additionally, the sensitivity of the spin-to-charge conversion is enhanced using a latched PSB mechanism \cite{Harvey-Collard2018, Zhao2019} by pulsing quickly from C to D.

The histogram shown in Fig.~\ref{fig:2}F is formed by running 30,000 single shot measurements and recording the difference of the SET current at pulse level D ($I_\text{D}$) and a reference level in the (4,0) region ($I_\text{R}$). The histogram reveals two peaks corresponding to measurements of singlet or triplet states. The singlet and triplet peaks are separated by 8.2\,$\sigma$, indicating a charge state readout infidelity on the order of $10^{-16}$. Since the pulse sequence A-D ideally initializes and measures the system in the $\ket{\downarrow\downarrow}$ state, the appearance of the singlet histogram peak allows us to infer a $\ket{\downarrow\downarrow}$ initialization fidelity of 81\%.

\paragraph*{Single-Electron Spin Resonance in a Global Field.}\label{section::ESR}

\begin{figure}[h!]
	\centering
	\includegraphics[]{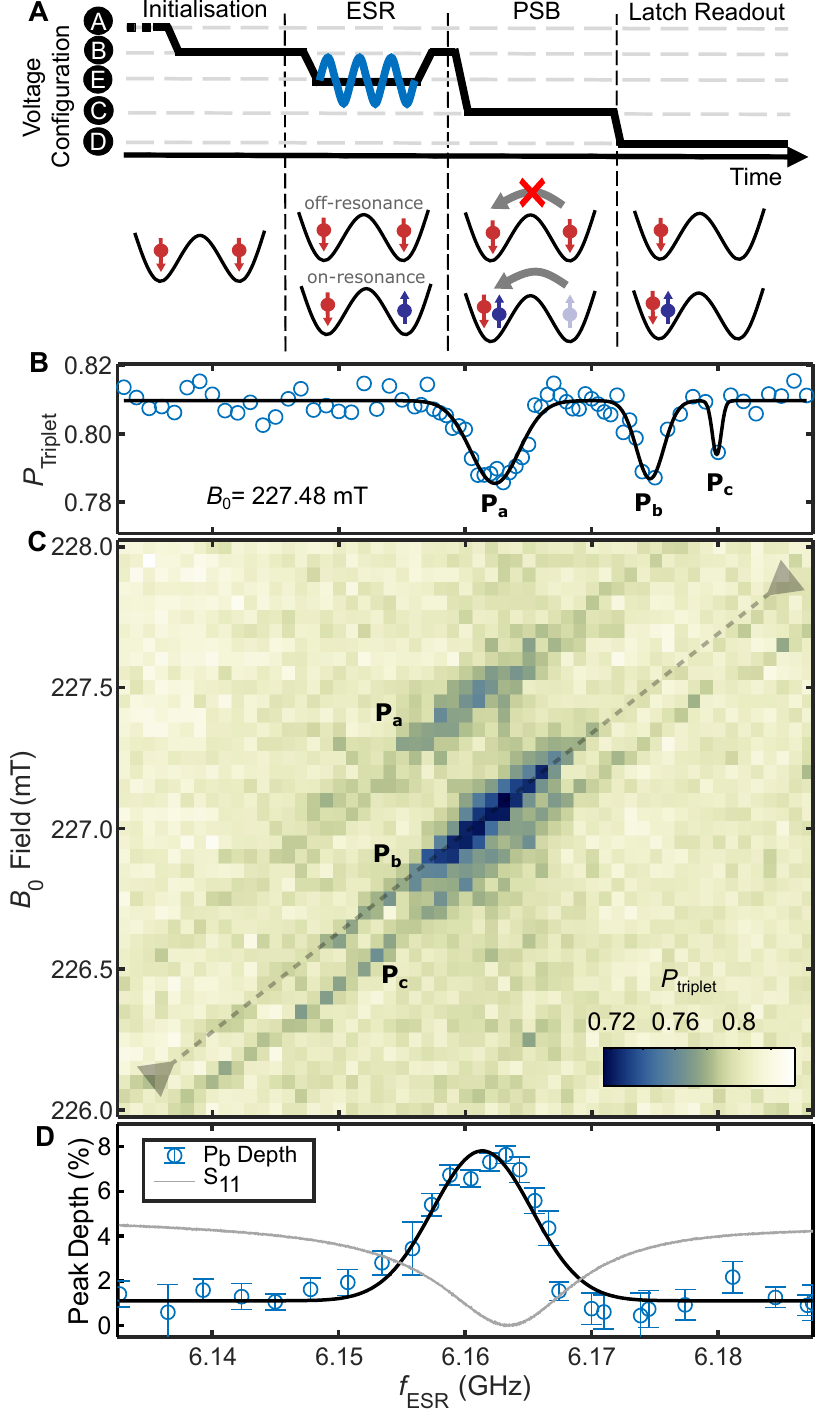}
	\caption{\textbf{Electron Spin Resonance Results.}
		\textbf{(A)} Pulsing scheme for the electron spin resonance measurements. The double quantum dot is initialized in a $\ket{\downarrow\downarrow}$ state. Microwave power is then applied to the dielectric resonator, generating an alternating magnetic field, $B_1$, which can rotate the spins if they are in resonance with the field. Finally, readout is performed to find the probability for the system to be in the triplet state. The cartoon below illustrates the spin states of the double dot system for each pulse stage, showing both the case when the ESR drive is on-resonance and when it is off-resonance.
		\textbf{(B)} Triplet probability as a function of the applied microwave frequency $f_{\rm ESR}$ at $B_0 = 227.48$\,mT, showing three ESR (electron spin resonance) peaks, two of which are consistent with the double dot system, while the third is an unexpected ESR peak. Note that panels B-D all share the same horizontal axis.
		\textbf{(C)} Triplet probability as a function of $f_{\rm ESR}$ and $B_0$, demonstrating that the ESR peaks shift with magnetic field, as expected.
		\textbf{(D)} Depth of the middle ESR peak (P\textsubscript{b}) plotted as a function of the qubit frequency, demonstrating enhancement of the magnetic field inside the bandwidth of the DR. The resonator $ S_{11} $ is overlaid in gray (see Fig.~\ref{fig:1}G). 
	}
	\label{fig:3}
\end{figure}

Having demonstrated spin initialization and readout, we now investigate electron spin resonance in a global microwave magnetic field. Using the pulse sequence illustrated in Fig.~\ref{fig:3}A, we first initialize a $\ket{\downarrow\downarrow}$ state in the double QD system. We then apply microwave control pulses to the coaxial loop coupler which excites the TE$_{11\delta}$ mode of the dielectric resonator, generating a global alternating magnetic field, $ B_{1} $, to manipulate the individual spins. When the DR frequency is resonant with either the $\ket{\downarrow\downarrow}$$\,\Longleftrightarrow\,$$\ket{\uparrow\downarrow}$ or $\ket{\downarrow\downarrow}$$\,\Longleftrightarrow\,$$\ket{\downarrow\uparrow}$ transitions (see Fig.~\ref{fig:2}D), which occurs at specific values of $B_{0}$, the spin states become mixed, reducing the probability of the system being in the $\ket{\downarrow\downarrow}$ state. The resonance frequencies can be calculated with $ f_{\rm res} =  g\mu_\text{B}B_{0}/h$, where $g$, $\mu_\text{B}$, and $h$ are the electron $g$-factor, Bohr magneton, and Planck constant, respectively. After spin manipulation, readout is performed by pulsing to the latched region (as described above) and classifying the detected spin state as either triplet or singlet depending on the size of the recorded SET current. Repeating this sequence several times allows us to calculate the probability for measuring a $\ket{\downarrow\downarrow}$ state $P_\text{triplet}$. When the spins in either Dot 1 or Dot 2 are resonant with the DR, we measure a reduction in $P_\text{triplet}$. It is known that the electron $g$-factors can be different in each dot due to electric-field-induced Stark shifts and device strains produced by thermal contraction of the metal gates \cite{Huang2017,Tanttu2019}.

Figure~\ref{fig:3}B shows the ESR peaks driven by the DR as a function of applied microwave frequency at a static magnetic field $ B_0 = 227.48$\,mT. By fitting the experimental data (blue circles) to Gaussian distributions we obtain two broad peaks P\textsubscript{a} and P\textsubscript{b}, with resonance frequencies of 6.163\,GHz and 6.174\,GHz, respectively, attributed to the double QD system, as well as a third peak P\textsubscript{c} at 6.180\,GHz, which could potentially result from an unintended spin state coupled to the DQD. To demonstrate that these peaks are spin-related, we measure the triplet probability as a function of $B_0$ and the $B_1$ drive frequency $f_{\rm ESR}$, as shown in Fig.~\ref{fig:3}C. All three peaks exhibit a linear dependence on $B_0$, each having $\approx54\,\text{MHz}$ shift in a range of 2\,mT $ B_0 $ field. The ESR frequencies are all consistent with electron spin g-factors in SiMOS quantum dots \cite{Tanttu2019}, which are generally slightly below $ g=2 $, to an accuracy limited by the calibration of our DC superconducting magnet.

We observe that the visibility of the P\textsubscript{a} and P\textsubscript{b} signals is enhanced considerably when the microwave frequency $f_{\rm ESR}$ is within the bandwidth of the DR fundamental mode, as should be the case if ESR is being driven by a global field produced by the resonator. In Fig.~\ref{fig:3}D, we plot the ESR peak depth along the P\textsubscript{b} transition as a function of ESR frequency, $ f_{\rm ESR} $, indicated by the dashed line in Fig.~\ref{fig:3}C and fit the result with a Lorentzian curve (black line). It is clear that the maximum peak depth coincides well with the $S_{11}$ response of the DR (Fig.~\ref{fig:1}G) re-plotted with grey line in Fig. \ref{fig:3}D for comparison). Outside of the DR resonance bandwidth, the visibility of P\textsubscript{b} saturates to approximately 1\,\% indicating that there is some residual drive which may, for example, originate from the coaxial loop coupler. However, the visibility is improved to 8\,\% at the resonance frequency (6.163\,GHz), demonstrating that the observed ESR is largely driven by the dielectric resonator itself. For powers exceeding $-32$\,dBm at the loop coupler, we observe transients in the measured SET current after application of the ESR pulse (see Methods), which coincides with the spin states being left in a fully mixed state. Consequently, we are not able to increase the microwave drive power beyond the ESR linewidth ($\Delta f = 2\,\text{MHz}~\text{to}~4\,\text{MHz}$) of the spins in this \textsuperscript{nat}Si device, which is a prerequisite for observing Rabi oscillations. 

\section*{Discussion}
A natural next step is the demonstration of coherent spin control. The measured device is fabricated on a \textsuperscript{nat}Si substrate which consists of approximately 4.7\,\% \textsuperscript{29}Si, having non-zero nuclear spins that cause significant inhomogeneous broadening of the resonance frequencies. Coherent spin driving therefore requires relatively fast Rabi frequencies, at least as large as the P\textsubscript{a} and P\textsubscript{b} transition linewidths of 4\,MHz and 2\,MHz, respectively (Fig. \ref{fig:3}C). Moving to an isotopically-enriched silicon substrate will substantially reduce the power requirements for observing coherent control \cite{Veldhorst2014}.

The measured internal quality factor of the dielectric resonator and device assembly is approximately two orders of magnitude lower than the material limit for KTO ($Q_\text{i}\approx 60,000$).  Improvements in the quality factor could be made through device modifications to remove any microwave current loops and to minimize the overlap of the DR with lossy materials such as bond wires and the highly-doped source and drain n$^+$ ohmic contacts. Reaching the material limit for $Q$ would boost the conversion factor by an order of magnitude, decreasing the power requirements a hundredfold and allowing Rabi frequencies as large as 6\,MHz for an input microwave power of just 200\,\textmu W.

The successful demonstration of electron spin resonance in a nanoelectronic device using a global magnetic field, as reported here, is a crucial step on the path to scale-up of spin-based quantum processors. This work shows that 3D dielectric resonators can be integrated with nanoelectronic circuits and are a viable source of global magnetic fields. The surface area of the current dielectric resonator ($0.7 \times 0.55$\,mm$^2$) would overlap with approximately 40 million qubits, assuming a conservative 100\,nm qubit pitch \cite{Veldhorst2017}. This area can be readily increased by reducing the height and/or frequency of the dielectric resonator. We believe that with the proposed alterations to the chip design and resonator assembly, large-scale control of millions of spin qubits in a continuous microwave field is now a realistic prospect.

\section*{Materials and Methods}

\paragraph*{Electromagnetic simulations.}
	To estimate the dielectric resonator conversion factor $C$ and plot the magnetic and electric field profiles in Fig.~\ref{fig:1} (E and F), we use the software package
	Circuit Simulation Technology Microwave Studio (CST-MWS) \cite{CST}. We define a 3D model of the global control device stack (using a simplified nanoelectronic device model), a coaxial loop coupler and copper enclosure. A waveguide port is defined to couple the microwave excitation to the coaxial loop and the frequency domain solver is used to solve Maxwell's equations over
	a finite-element mesh of the model.
	
	The internal quality factor observed in the experiment is reproduced in the simulation by adjusting the loss tangent of materials in the device and the external coupling strength is matched by adjusting the position of the loop coupler. The magnetic and electric field profiles are extracted by placing a field monitor at the resonant frequency, which is found by observing the simulated $S_{11}$ parameter response. To estimate the magnetic field homogeneity of the DR we simulate the KTO and silicon (without the device layer), using the eigenmode solver to calculate the magnetic field distribution of the TE$_{11\delta}$ mode. This is presented in Fig.~\ref{fig:Methodshomogeneity} as a percentage of the maximum field value. 

\begin{metfig}[!h]
	\centering
	\includegraphics[]{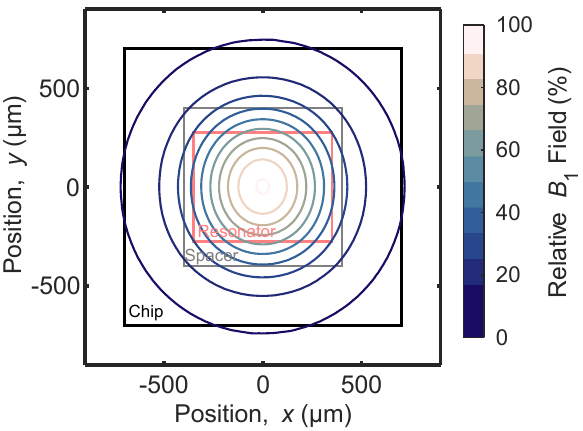}
	\caption{\textbf{Magnetic field homogeneity.}
		Magnetic field distribution of the TE$_{11\delta}$ mode, plotted as a percentage of the maximum field value at the surface of the silicon sample.
	}
	\label{fig:Methodshomogeneity}
\end{metfig}

\paragraph*{Experimental set-up.}
	Our experiments were performed in a set-up similar to previous work \cite{Veldhorst2014,Yang2020} except for the delivery of the microwave signal to the device. The diagram in Fig.~\ref{fig:MethodsSetup} depicts how the signal generated by the microwave source is routed to the device. The signal reaches the coaxial loop coupler after going through a cryogenic circulator with a pass-band of 4\,GHz to 8\,GHz. The microwave signal produces an alternating current around the loop, generating a magnetic field that inductively couples to the DR field, as explained in the section `\nameref{section::DR}'. To measure the DR $S_{11}$ response we replace the microwave source with a vector network analyzer (VNA), where the reflected signal from the loop coupler is returned to the VNA via the circulator.

\begin{metfig}[!h]
	\centering
	\includegraphics[]{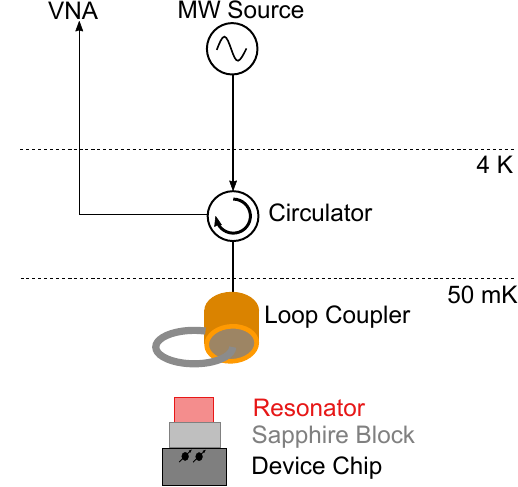}
	\caption{\textbf{Microwave control setup diagram.} 
		Schematic showing the connections between the microwave (MW) source, VNA used for the $S_{11}$ measurements and coaxial loop coupler.
	}
	\label{fig:MethodsSetup}
\end{metfig}

\paragraph*{SET response to a continuous microwave drive.}
	Whilst the ESR experiments of section `\nameref{section::ESR}' are performed in pulsed mode, where microwaves are applied in 150\,\textmu s bursts (a duty cycle of approximately 0.7\,\%), here we present the SET current response under a continuous microwave drive at the DR fundamental mode resonance frequency. We scan the SET ST gate voltage ($V_{\rm ST}$) and monitor the current $I_{\rm SET}$ as we increase the power of the microwave drive (Fig. \ref{fig:MethodsCP}). The broadening of the observed Coulomb peak widths can be a useful measure to investigate the effect of electric fields present in the device whilst driving. For powers starting from $-38$\,dBm, the broadening linearly increases as a function of power. We note that even with powers as high as $-25$\,dBm at the coupler, there is still a good visibility of the peaks.

\begin{metfig}[!h]
	\centering
	\includegraphics[]{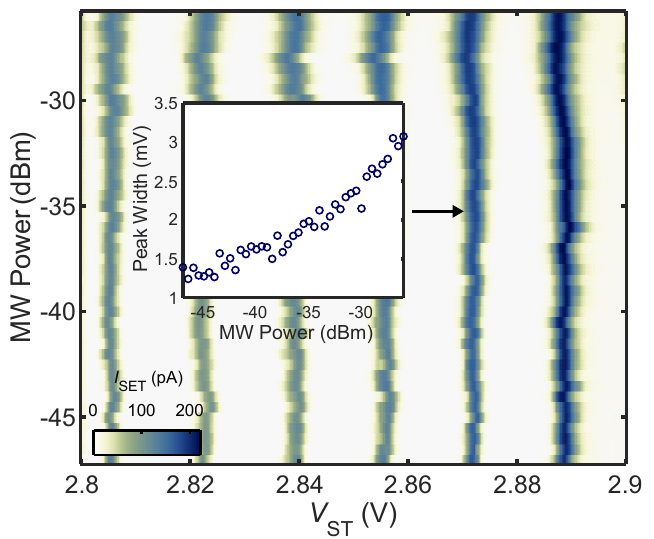}
	\caption{\textbf{Coulomb peaks versus microwave power.} 
		Coulomb peaks measured as a function of the microwave power applied to the loop coupler. Inset: Width of the Coulomb peak identified by the black arrow.
	}
	\label{fig:MethodsCP}
\end{metfig}

\paragraph*{Pauli spin blockade search.}
	The 2D color-map in Fig.~\ref{fig:2}E of the main text is a measurement primarily performed to search for the Pauli Spin-Blockade (PSB) region at the (4,0)-(3,1) transition. Each pixel in the map encodes the current difference $\Delta I_{\rm SET}$ between two measurements with different spin initializations: A mixed state (producing current $I_{\rm mix}$) and a singlet state ($I_{\rm singlet}$). Both types of initializations are depicted in Fig. \ref{fig:MethodsPSB}. The current $I_\text{mix}$ is measured by following the sequence M1-M2-R, which starts with (3,0) and then ramps to (3,1) before readout. The spin state of the electron loaded to D2 after the ramp could be parallel or anti-parallel to the spin of the third electron in D1, i.e. the two spin-state becomes a mixture of singlet and triplet states. For measuring $I_\text{singlet}$ the sequence becomes S1-S2-R. The resulting charge configuration after the ramp to S2 becomes (4,0), which forces the two-spin state to be a singlet due to PSB. The voltages of points M1, M2, S1 and S2 are fixed in the experiment whilst R is scanned, with the ``readout search signal'' $\Delta I_{\rm SET} = I_{\rm mix} - I_{\rm singlet}$ measured at each pixel to produce the color map.

\begin{metfig}[!h]
	\centering
	\includegraphics[]{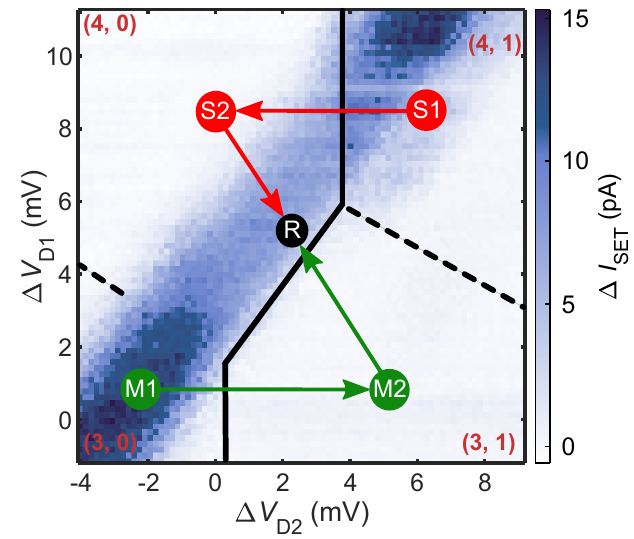}
	\caption{\textbf{Pauli Spin Blockade Search Measurement.} 
		Green (red) level labels and arrows indicate how mixed (singlet) states are initialized for PSB search. The readout search signal $\Delta I_{\rm SET} = I_{\rm mix} - I_{\rm singlet}$ is measured at the level R, which is swept all over the 2D space in the plot.
	}
	\label{fig:MethodsPSB}
\end{metfig}

\paragraph*{Power limit to ESR pulses.}
	As mentioned in the section `\nameref{section::ESR}', ESR pulses with powers exceeding $-32$\,dBm at the loop coupler produced random charge jumps in the SET current (Fig.~\ref{fig:MethodsSpinScrambling}). These jumps were observed to completely mix the spin state of the system, i.e. remove the spin information before read-out. For lower powers, the traces do not show any charge jumps and the read-out was not affected.   

\begin{metfig}[!h]
	\centering
	\includegraphics[]{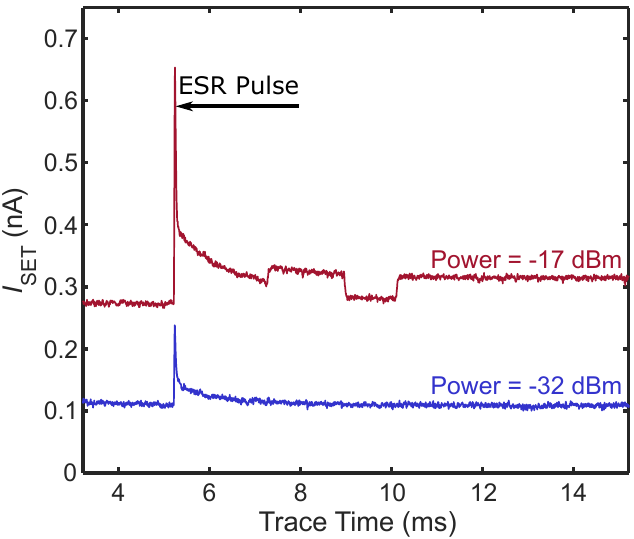}
	\caption{\textbf{Time-resolved SET current traces recorded for ESR pulses with different powers.} 
		The trace with high power ($-17$\,dBm at the coupler) has unexpected steps in the SET current, indicating sudden electron jumps in the vicinity of the dots and SET. The jumps are not observed in the low power trace ($-32$\,dBm at the coupler). Traces offset by a 0.15\,nA vertical shift.
	}
	\label{fig:MethodsSpinScrambling}
\end{metfig}

\paragraph*{Background subtraction.} The data in Fig.~\ref{fig:3}B and 2D color-map in Fig.~\ref{fig:3}C are post-processed to remove background artefacts unrelated to the spins. A horizontal trace recorded at a $B_0$ far off resonance with spins is subtracted from the data. Additionally, each horizontal trace is offset by its own median current to correct for any variation in the SET current level over time.



\bibliography{OffChipReferences}
\bibliographystyle{naturemag}

\section*{Acknowledgments}
\paragraph*{Funding:} The authors acknowledge support from the Australian Research Council (DE190101397, FL190100167 and CE170100012), the US Army Research Office (W911NF-17-1-0198) and the NSW Node of the Australian National Fabrication Facility. The views and conclusions contained in this document are those of the authors and should not be interpreted as representing the official policies, either expressed or implied, of the Army Research Office or the US Government. E.V. and J.P.S.-S. acknowledge support from Sydney Quantum Academy.
\paragraph*{Author contributions:}
	E.V. and J.P.S.-S. performed the experiments. J.P.S.-S. and J.J.P. designed and fabricated the DR and J.P.S.-S. ran the electromagnetic simulations. T.D. designed the copper sample enclosure and assisted with the DR characterization. W.H.L. and F.E.H. fabricated the silicon device. E.V., J.P.S.-S. and R.C.C.L. analyzed the data. C.H.Y., T.T. and A.L. contributed to discussions on the experimental results. E.V., J.P.S.-S., J.J.P., W.H.L. and A.S.D. wrote the manuscript with input from all authors. J.J.P. and A.S.D. supervised the project.
\paragraph*{Data and materials availability:}
	All data needed to evaluate the conclusions in the paper are present in the paper. Additional data related to this paper may be requested from the authors.

\end{document}